\ifdefined\USEIEEE

\documentclass[conference]{IEEEtran}

\IEEEoverridecommandlockouts
\usepackage{cite}
\usepackage{amsmath,amssymb,amsfonts}
\usepackage{algorithmic}
\usepackage{graphicx}
\usepackage{textcomp}
\usepackage{xcolor}
\usepackage{booktabs}

\def\BibTeX{{\rm B\kern-.05em{\sc i\kern-.025em b}\kern-.08em
    T\kern-.1667em\lower.7ex\hbox{E}\kern-.125emX}}

\else 
    \documentclass[runningheads]{llncs}
    \usepackage{graphicx}
    \usepackage{booktabs}
    \usepackage{cite}
    \usepackage{amsmath,amssymb,amsfonts}
    \usepackage{algorithmic}
    \usepackage{textcomp}
    \usepackage{xcolor}
    \usepackage{booktabs}
\fi

\usepackage{lipsum,lineno}
\begin{document}
\title{Educational Robotics in Online Distance Learning: An Experience from Primary School}
\titlerunning{Educational Robotics in Online Distance Learning for Primary School}
%
\author{Christian Giang\inst{1,2}\orcidID{0000-0003-2034-9253}
\and
Lucio Negrini\inst{1}\orcidID{0000-0001-6793-6258}
}
\authorrunning{C. Giang and L. Negrini}
%
\institute{Department of Education and Learning, University of Applied Sciences and Arts of Southern Switzerland (SUPSI), Switzerland \and Ecole Polytechnique Fédérale de Lausanne (EPFL), Switzerland
\\
\email{firstname.lastname@supsi.ch}}

\maketitle

\begin{abstract}
Temporary school closures caused by the Covid-19 pandemic have posed new challenges for many teachers and students worldwide. Especially the abrupt shift to online distance learning posed many obstacles to be overcome and it particularly complicated the implementation of Educational Robotics activities. Such activities usually comprise a variety of different learning artifacts, which were not accessible to many students during the period of school closure. Moreover, online distance learning considerably limits the possibilities for students to interact with their peers and teachers. In an attempt to address these issues, this work presents the development of an Educational Robotics activity particularly conceived for online distance learning in primary school. The devised activities are based on pen and paper approaches that are complemented by commonly used social media to facilitate communication and collaboration. They were proposed to 13 students, as a way to continue ER activities in online distance learning over the time period of four weeks.



\ifdefined\USEIEEE
\else 
\fi 
\end{abstract}

\section{Introduction}

With the onset of the Covid-19 pandemic, schools worldwide faced temporary closure, intermittently affecting more than 90\% of the world’s students \cite{donohue2020covid}. As a consequence, many teachers had to adapt their activities to online formats in order to continue instruction at distance. Moving to the online space represented a particular challenge for compulsory schools, where, in contrast to many universities, online learning is still rather unexplored \cite{allen2020teaching}. In addition to general difficulties associated with this new format of instruction, such as lack of experience in online learning or poor digital infrastructure\cite{carrillo2020covid}, implementing Educational Robotics (ER) activities appeared to be particularly complicated. ER activities usually comprise a combination of different learning artifacts \cite{giang2019heuristics}, namely robots, programming interfaces and playgrounds, that are often dependent from each other and have thus also been referred to as "Educational Robotics Learning Systems" (ERLS) \cite{giang_towards_2020}. Many times, students benefit from their school’s infrastructure providing them with access to these artifacts. In this regard, the restrictions on individual mobility imposed during the pandemic in many countries, have significantly limited these opportunities. Although previous works have explored remote labs as a way to implement robotics education in distance learning, most approaches were limited to university level (e.g. \cite{kulich2012syrotek,dos2016web,di2017adaptive,farias2019development}) or secondary school (e.g. \cite{despres1999computer,almeida2017remote}). Bringing such approaches to primary school education is not obvious, since it would require extensive adjustments of the tools and activities proposed to make age-appropriate adaptations.

Furthermore, ER activities usually capitalize on students’ interactions with their peers and teachers building on the ideas of project-based didactics \cite{kilpatrick1926project} and social constructivist learning \cite{vygotsky1980mind}. As illustrated in a previous study by Negrini and Giang \cite{negrini2019pupils}, students indeed perceived collaboration as one of the most important aspects of ER activities. However, during the pandemic, with teachers and students working from home, providing possibilities for direct interactions became significantly more difficult. While the recent technological advances have yielded a myriad of tools for online communication and collaboration, only few educators have already acquired the pedagogical content knowledge to leverage them for the design of meaningful online learning experiences \cite{rapanta2020online}. As a matter of fact, Bernard et al. \cite{bernard2004does} have emphasized that the effectiveness of distance education depends on the pedagogical design rather than the properties of the digital media used. Designing engaging and pedagogically meaningful online activities is especially important at primary school level, in order to address the limited online attention spans of young children as well as the concerns of many parents with regard to online learning \cite{dong2020young}. 

Both the limited access to ER learning artifacts as well as the unfavorable conditions for direct interactions in online settings, thus represented major obstacles to be overcome to implement ER online activities during the Covid-19 pandemic. Yet on the other hand, these circumstances emerged as a possibility to experiment with new approaches addressing these issues. In this work, we will present one such approach made during the Covid-19 pandemic to implement ER online activities for primary school students. To address the issue of limited access to ER learning artifacts, activities were mainly based on pen and paper approaches. Commonly used social media were used as a complement to augment the learning experience and to allow students to communicate with their peers and teacher as well as to collaborate with each other. The devised activities were proposed to 13 students aged between 9 and 12 years old, over the time course of four weeks. In the following, we will present the devised activity, report on the benefits and challenges observed from this experience and finally discuss how the findings could inform future works in this domain.

\section{An example of an ER online activity for primary school} 
\label{sec:Development}

\begin{figure}
    \centering
    \vspace{-10pt}
    \includegraphics[width=\linewidth]{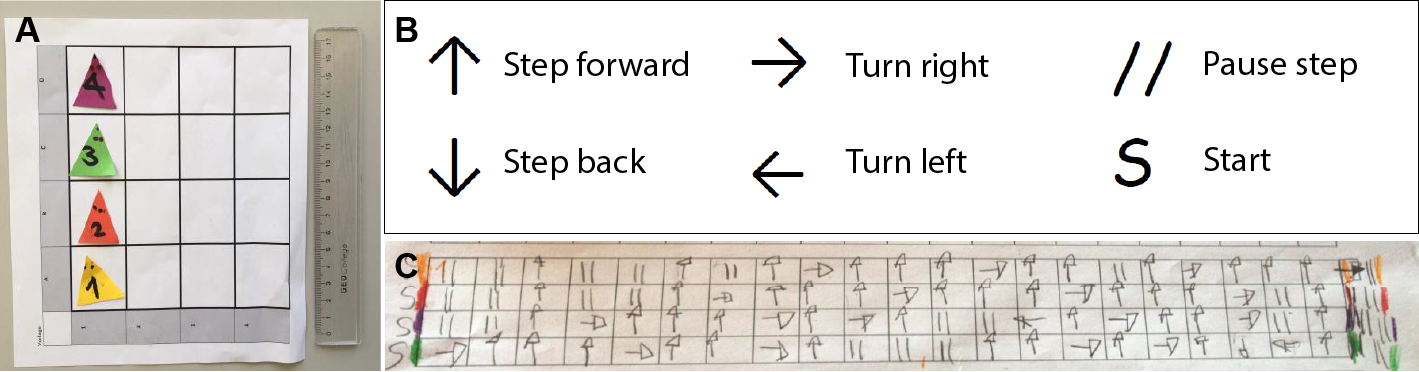}
    \vspace{-10pt}
    \caption{Paper grid and triangles to locally simulate the behavior of the Bee-Bots (A), set of instructions for the robot (B) and an example of a solution proposal (C).}
    \vspace{-15pt}
    \label{fig:material}
\end{figure}

The devised ER online activities were proposed to 13 primary school students from a small village in Switzerland (aged between 9 and 12 years old). In the German-speaking part of Switzerland the compulsory school curriculum foresees “Media and informatics” activities but leaves it to the teacher to decide which tools they want to use for those activities. In this case, the teacher decided to perform ER activities during those classes as well as during math classes. Therefore, in the classroom activities performed before the pandemic, the students had the possibility to familiarize themselves with the Bee-Bot\footnote{https://www.tts-international.com/bee-bot-programmable-floor-robot/1015268.html} robot and its programming instructions. After school closure due to the pandemic, online activities were developed and proposed to the students as a voluntary option to continue the ER activities from home. From the 13 students that were offered this option, 9 decided to complete them. School lessons during the closure were organized on a weekly basis. At the beginning of each week, new assignments and learning materials were deployed to the students by email or by post, which were then discussed on the following day through online video conferences. The rest of the week was dedicated to the independent completion of the assignments with the possibility for individual discussions with the teacher. Since students did not have access to the necessary ER tools, a pen and paper approach was taken to provide them with alternative learning artifacts: colored triangles were used as robots, paper strips as programming interfaces and a paper grid as the playground (Fig. \ref{fig:material}). 

\begin{figure}
    \centering
    \vspace{-10pt}
    \includegraphics[width=\linewidth]{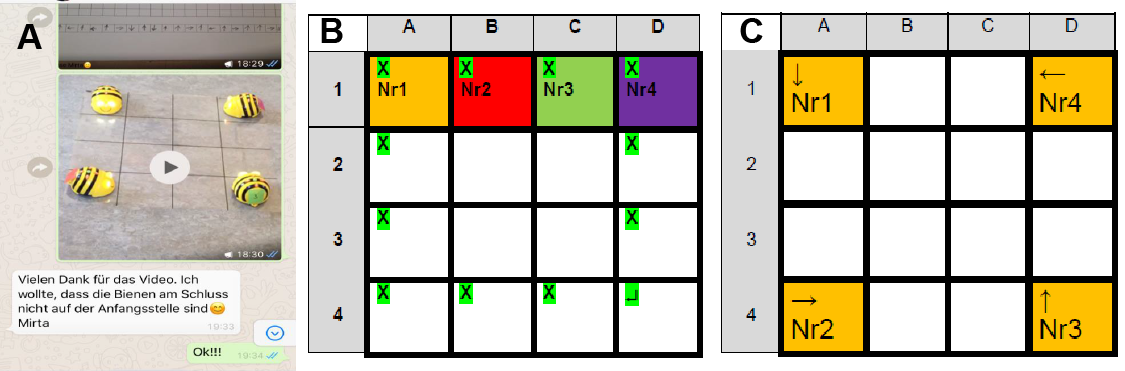}
    \vspace{-10pt}
    \caption{Example of a message exchange between a student and the teacher (A), assignment in which four robots have to be programmed to drive around the outer circle of the grid (B) and assignment in which all four robots have to perform the same choreography without crashing into each other (C).}
    \vspace{-15pt}
    \label{fig:activities}
\end{figure}

The paper grid and the triangles served the students as learning artifacts to locally simulate the behavior of the Bee-Bots. Each triangle represented a robot with its orientation on the grid. The paper strips allowed the students to write down their solution proposals before submitting them to their teacher for evaluation. Solutions were submitted as photos sent through smartphone messaging applications (Fig. \ref{fig:activities}A). The teacher, having access to Bee-Bot robots, programmed the robots according to the solution proposals submitted by the students and made video recordings of the resulting behaviors. The videos were then returned to the students together with some feedback of the teacher (using text or audio messages). Moreover, students could directly provide feedback to their peers, either during the synchronous video conferencing sessions or through exchanges in group chats. In total, 12 different tasks were developed, from which the students could choose one at the beginning of each week. To complete the tasks, students had to find a suitable series of programming instructions for the Bee-Bots that would satisfy certain conditions. For instance, in the activity illustrated in Fig. \ref{fig:activities}B, four robots (starting positions marked in colors) have to be programmed to drive around the outer circle of the grid (marked with crosses). The idea of this activity was for students to realize that a turn takes an additional step and they therefore had to make use of the ``pause step" command to successfully complete the task without making the robots crash into each other. In another example, illustrated in Fig. \ref{fig:activities}C, students had to program four Bee-Bots with the exact same programming instructions, making them perform a choreography and finally return to their starting positions (orange fields). The difficulty here was to find the longest possible choreography, without making robots crash into each other. The other ten activities were similar, with an increasing difficulty given by adding more robots, larger grids or more constraints. For all proposed activities, students worked independently from home. In case of difficulties they could contact the teacher that was available via social media.

\section{Results and Discussion}

The students who participated in the activities showed high engagement and evaluated the ER online activity as mainly positive. Students were usually eagerly awaiting the video recordings of their programmed Bee-Bot behaviors. Once they received the videos from their teacher, students uploaded them in the group chat, allowing others to provide feedback and suggest improvements. The integration of social media that most students were familiar with from their everyday life (i.e., smartphone messaging applications) facilitated a smooth operation without any technical issues. At the same time, it allowed students to communicate and collaborate with their peers even from afar. Similarly, it allowed the teacher to provide direct feedback to individual students using text or audio messages. In addition to the asynchronous approach through messaging applications, the weekly online video conferences enabled synchronous moments of interaction. During the video conferences, the teacher discussed with the students in plenary sessions, explaining upcoming assignments and debriefing completed ones. In their work published during the pandemic, Rapanta et al. \cite{rapanta2020online} have suggested mixing synchronous and asynchronous approaches as a favorable approach to online learning. When well designed, it allows to shift the responsibility for learning to the student, while the teacher takes the role of a facilitator and tutor. Since ER is built on the same learner-centered approach, following this idea of hybrid forms of instruction, could provide a favorable framework for the implementation of ER online activities. Moreover, from the teacher’s perspective, following hybrid approaches could significantly reduce the fatigue associated with the extensive use of synchronous forms of online instruction.  Except for the first week, where around 30 minutes were taken to explain the general idea of the devised activities, only little amounts of time (around 5 minutes) were needed to discuss the activities during the online video conferences of the subsequent weeks.

In a recent work, El-Hamamsy et al. \cite{el-hamamsy_computer_2020} have highlighted that the introduction of screens in primary school is still met with reticence. Reducing active screen time was particularly relevant during the pandemic, in order to address the limited online attention span of young children as well as the concerns of their parents with respect to online learning \cite{dong2020young}. With online learning becoming increasingly prevalent, finding alternative, “analogue” ways of distance learning is crucial. As Mehrotra et al. \cite{mehrotra_introducing_2020} have suggested before, one way to address the issue of extensive screen time in ER activities, is to introduce paper-based learning approaches. Building on this idea, the activities proposed in this work mainly relied on paper-based materials. It could be argued that performing ER activities without having students directly interact with robots and programming interfaces represents a major limitation of the proposed approach. However, a recent work by Chevalier et al. \cite{chevalier2020fostering} has also highlighted that providing students with unregulated access to ER tools, in particular the programming interfaces, does not necessarily result in better learning experiences. Indeed, when ER activities are not well designed, involving these learning artifacts can even be counterproductive, since they can promote blind trial-and-error approaches and hence prevent desired reflection processes. The pen and paper approach applied in this work provided students with alternative learning artifacts, that in contrast to real robots, cannot provide immediate feedback on the proposed solutions. The decelerated feedback loop, requiring students to take pictures, sending them to their teacher and waiting for their response, may therefore represent an interesting approach to promote the use of pauses for reflection. As shown by Perez et al. \cite{perez2017identifying}, the strategic use of pauses was strongly associated with successful learning in inquiry-based learning activities. 

The three youngest students (9 years old) who chose to participate in the proposed ER online activities, completed one assignment per week for the whole duration of the school closure (in total four weeks). From the six older students (12 years old) who started in the first week, only one half continued until the end. The higher dropout rate of the older students could be related to the lack of difficulty with regard to the proposed activities. Bee-Bots are designed as tools for rather young children and are therefore limited in the complexity of the possible programming instructions. The proposed ER online activities could thus have been perceived as not sufficiently challenging by the older students.

\section{Conclusion}
The presented ER online activity was developed to face the challenges of online distance learning and allowed primary school students to continue with ER activities during the Covid-19 pandemic. However, some interesting aspects emerged from this experience, in particular the promotion of reflection processes due to the use of pen and paper learning artifacts. The relevance of such approaches could indeed go beyond the context of online learning and potentially also generalize to ER in face-to-face classroom activities. However, it should be acknowledged that more exhaustive studies are needed to confirm this hypothesis. Moreover, future work could also investigate whether similar approaches could be adapted for more advanced ER tools in order to design more complex tasks for more mature students.

\section{Acknowledgements}
The authors would like to thank the teacher Claudio Giovanoli for his contributions in designing and implementing the Bee-Bot online activity as well as the children of the school in Maloja who decided to participate in the voluntary ER online activities.


\bibliography{biblio.bib}

\end{document}